\documentclass[twocolumn,eqsecnum,aps,prd]{revtex4}
\usepackage{amssymb}
\usepackage{bm}
\begin{document}
\begin{titlepage}

\title{QCD strings with spinning quarks}
         
\author{Theodore J. Allen} 
\affiliation{Department of Physics, Hobart and William Smith Colleges \\
Geneva, New York 14456 USA }

\author{M. G. Olsson}
\affiliation{Department of Physics, University of Wisconsin \\
1150 University Avenue, Madison, Wisconsin 53706 USA}

\author{Jeffrey R. Schmidt} 
\affiliation{Department of Physics, University of Wisconsin-Parkside\\ 
900 Wood Road, Kenosha, Wisconsin 53141 USA}

\date{\today}
\thispagestyle{empty}
\begin{abstract}
We construct a consistent action for a massive spinning quark on the
end of a QCD string that leads to pure Thomas precession of the
quark's spin.  The string action is modified by the addition of
Grassmann degrees of freedom to the string such that the equations of
motion for the quark spin follow from boundary conditions, just as do
those for the quark's position.
\end{abstract}
\maketitle

\end{titlepage}

\section{Introduction}\label{sec:intro}

A consistent description of spin within a QCD string theory has been
sought for many years.  The addition of dynamical spin to the bosonic
string led to the development of supersymmetry and superstring theory
\cite{Green:sp}. Such theories are more realistic as unified theories
of elementary particle physics than as phenomenological descriptions
of hadronic states.

A more realistic description of hadronic states involves the
replacement of the free end of the dual resonance string by the
addition of a massive point quark to the end of the string. In 1977
Ida \cite{Ida:1977uy} analyzed the motion of a spinless massive quark
on the end of a bosonic string.  The relativistic flux tube model
\cite{rft}, derived from different assumptions, is mathematically
equivalent to a bosonic string with a spinless quark end and produces
realistic meson spectra on average, but there is no place for quark
spin in this model.  In this paper we make a modification of the
bosonic string plus bosonic quark model to introduce quark spin.

Our clue to constructing a consistent action comes from the suggestion of
Buchm\"uller \cite{Buchmuller:1981fr} that the spin of the quark should
undergo pure Thomas precession because the quark sees a purely
chromoelectric field in its rest frame.  This seems to be supported by
experimental data \cite{so,Isgur:kr} and is in agreement with QCD
\cite{eichten,Bali:1997am}. 

We begin in Sec.\ \ref{sec:pseudo} by discussing the treatment of spin
in pseudoclassical language.  We show how to construct actions for a
free fermion as well as a fermion with background scalar and vector
potentials.  We analyze the case of a scalar potential in detail and
show how the Thomas precession manifests itself in this language.

In Sec.\ \ref{sec:thomas} we show in detail that the Fermi-Walker
transport of the spin vector, which is the equation of motion of the
spin vector for a particle in a scalar potential, leads to Thomas
precession of the spin in its rest frame.

In Sec.\ \ref{sec:fixedstring} we use the example of a spinless quark
coupled to the end of a string to argue for the form of the action for
a spinning particle coupled to a modified Polyakov string action.  The
key idea is to obtain the equations of motion of the spin of the quark
from boundary conditions, just as the equations of motion of the
quark's position arise from boundary conditions.  To this end, we
introduce new Grassmann-valued fields on the string worldsheet.

In Sec.\ \ref{sec:alphabeta} we use the consistency of the equations
of motion of the quark and the requirement of Thomas precession to fix
the parameters in the string action.  The result is that the only
modification of a free spinning quark plus free bosonic string action
is the replacement of the bosonic string position variable by the 
string position variable plus a term bilinear in worldsheet fermionic
variables.

In Sec.\ \ref{sec:fermiongauge} we explore the fermionic gauge invariance
of our string action. In the phenomenologically interesting case, we
find that the worldsheet fermionic variables are pure gauge degrees of
freedom. 

We find the momentum and angular momentum from Noether's theorem in
Sec.\ \ref{sec:noether}.  These conserved quantities are the usual
starting point for the numerical quantization of the relativistic flux
tube model.  Finally, we conclude in Sec.\ \ref{sec:discussion}.

\section{Spin in pseudoclassical mechanics}\label{sec:pseudo}

We choose to work within the framework of pseudoclassical mechanics
\cite{ref:pseudoclassical} because the formalism is elegant as well as
physically transparent; the transition from pseudoclassical to quantum
mechanics is immediate.  In this section we construct actions that
produce the Dirac equation, both free and in background potentials, as
an equation of motion and we show how the Thomas precession in a
scalar potential manifests itself in this language.  The main
disadvantage is that it requires some familiarity with the technical
details of Dirac's constrained Hamiltonian mechanics
\cite{ref:Dirac,ref:Henneaux} as well as classical mechanics with
Grassmann variables \cite{ref:pseudoclassical,ref:Henneaux}.

The easiest way to construct pseudoclassical actions for fermions is
to consider the Dirac equation as a phase-space constraint and to
construct consistent actions that yield this constraint. The first
actions of this type were found by Berezin and Marinov \cite{Berezin},
Barducci, Casalbuoni, and Lusanna \cite{Casalbuoni}, and Brink, Deser,
Zumino, Di Vecchia, and Howe \cite{Brink}.  To represent the spin
degrees of freedom of a fermion, a set of five Grassmann coordinates,
$\xi_\mu$ and $\xi_5$, are introduced.  Upon quantization, the
Grassmann coordinates will become generators of a Clifford algebra and
can be identified with Dirac's gamma matrices.  The kinetic piece of
the action for the Grassmann variables
\begin{equation}
S_{\rm kinetic} = \int d\tau {i\over 2}\left(\xi_\mu \dot\xi^\mu + \xi_5
\dot\xi_5\right),
\end{equation}
leads to the canonical second-class constraints
\begin{equation}\label{eq:Chis}
\chi_\mu = \pi_\mu - {i\over2} \xi_\mu \approx 0 , \qquad \chi_5 = \pi_5 -
{i\over2} \xi_5 \approx 0 .
\end{equation}
Here we use Dirac's wavy equal sign notation
\cite{ref:Dirac,ref:Henneaux} for ``weak equality,'' which reminds us
that the equalities cannot be taken before Poisson brackets are
calculated.  We denote the canonical momenta to $\xi^\mu$ and $\xi_5$,
defined to be the derivative of the Lagrangian from the right with
respect to the velocities $\dot\xi^\mu$ and $\dot\xi_5$ respectively,
by $\pi_\mu$ and $\pi_5$ .  With this convention, we obtain the
following Poisson brackets
\begin{eqnarray}
\{\xi^\mu,\pi_\nu\}  & = & \{\pi_\nu, \xi^\mu\} = \delta_\nu^\mu , \\
\{\xi_5,\pi_5\} & = & \{\pi_5, \xi_5\} = 1,
\end{eqnarray}
with all others being zero. Our conventions for pseudoclassical
mechanics are given in Appendix \ref{sec:PCMC}.

The weak equalities in Eq.~(\ref{eq:Chis}) can be replaced by strong ones
if we introduce the Dirac brackets \cite{ref:Dirac}.  From the definition in
Appendix \ref{sec:DiracBrackets} and the Poisson brackets above, we find
\begin{eqnarray}
\{\xi_\mu,\xi_\nu\}_D & = & -i\eta_{\mu\nu} , \\
\{\xi_\mu, \xi_5\}_D & = & 0 , \\
\{\xi_5, \xi_5\}_D & = & -i ,
\end{eqnarray}
where $\eta_{\mu\nu}$ is the metric. Our convention is 
$\eta_{\mu\nu}={\rm diag}(-1,+1,+1,+1)$. 

The meaning of the Grassmann numbers becomes clear upon quantization.
When we make the replacement of $i\hbar$ times Dirac brackets by
anticommutators, we find that the quantum operators $\widehat\xi_\mu$
and $\widehat \xi_5$ obey a Clifford algebra,
\begin{eqnarray}\label{eq:anticom}
\widehat\xi_\mu \widehat\xi_\nu + \widehat\xi_\nu \widehat\xi_\mu & =
& \hbar\eta_{\mu\nu}, 
\nonumber \\
\widehat\xi_\mu \widehat\xi_5 + \widehat\xi_5 \widehat\xi_\mu & = & 0 , \\
\widehat\xi_5 \widehat\xi_5 & = & \frac{\hbar}2 . \nonumber
\end{eqnarray}
From these anticommutation relations, we see that the operators
$\widehat\xi_\mu$ and $\widehat\xi_5$ can be represented as gamma
matrices,
\begin{eqnarray}
\widehat\xi_\mu & = & \sqrt{\hbar\over 2} \gamma_5\gamma_\mu , \\
\widehat\xi_5 & = & \sqrt{\hbar\over 2} \gamma_5 .
\end{eqnarray}
The free Dirac equation is proportional to 
\begin{equation}\label{eq:freeDirac}
\widehat\phi | \psi \rangle = \left( \widehat p_\mu\widehat\xi^\mu +
m\widehat\xi_5 \right) | \psi \rangle = 0 .
\end{equation}
Thus, we should introduce the constraint
\begin{equation}
\phi = p_\mu \xi^\mu + m \xi_5 \approx 0 
\end{equation}
into our action.  This constraint does not have vanishing Dirac
bracket with itself, but yields the Klein-Gordon operator:
\begin{eqnarray}\label{eq:freeKG}
K & \equiv & \frac{i}2 \big\{p_\mu\xi^\mu + m\xi_5,\, p_\mu\xi^\mu +
m\xi_5 \big\}_D \nonumber \\ 
& = & \frac12 (p^2 + m^2) \approx 0 .
\end{eqnarray}
In order to be able to impose the constraint $\widehat\phi$ as in
Eq.~(\ref{eq:freeDirac}), $\phi$ and any constraints, such as $K$, arising
from it must be first-class, which means the Dirac brackets of any pair of
them yields a combination of other first-class constraints.  In order for
the set of constraints to close under Dirac brackets, this last constraint
must have vanishing Dirac bracket with $\phi$. This is guaranteed by the
(graded) Jacobi identity,
\begin{equation}
\{\phi, K\}_D = {i\over 2} \big\{\phi , \{\phi, \phi\}_D\big\}_D = 0 \ .
\end{equation}

The dynamics of this system are given by the free action, plus these
constraints put in with Lagrange multipliers $\lambda$, and $e$:
\begin{eqnarray}\label{eq:freeAction}
S & = & \int d\tau\, \Big[p_\mu \dot x^\mu + \frac{i}2(\xi_\mu \dot\xi^\mu +
\xi_5 \dot \xi_5)\qquad\phantom{.} \nonumber \\
& &\phantom{\int} +\, i{\lambda\over m} (p_\mu\xi^\mu + m\xi_5) - e
\frac12(p^2 + m^2)\Big] \ . 
\end{eqnarray}

We may eliminate $p$ from Eq.~(\ref{eq:freeAction}), by using its (purely
algebraic) equation of motion.  Similarly, we may then eliminate $e$ from
the intermediate action to find the action given by Berezin and Marinov
\cite{Berezin}
\begin{eqnarray}\label{eq:freefermion}
S & = & \int d\tau\, \Big[- m \sqrt{- \dot x^2} + \frac{i}2(\xi_\mu
\dot\xi^\mu + \xi_5\dot \xi_5)   \nonumber \\ 
 & & \phantom{\int d\tau\, \Big[} 
+\, i \lambda (\xi_5 + u\cdot\xi) \Big] \ ,
\end{eqnarray}
where we have used the usual notation for the four-velocity, $u = \dot x
/\sqrt{-\dot x^2}$.

The Dirac equation in a background scalar field $\varphi$ and vector
field $A_\mu$ is obtained from the free equation by minimal
substitution for $A_\mu$ and the addition of $\varphi$ to the mass:
\begin{equation}
\phi = (p - A)_\mu \xi^\mu + (m+\varphi) \xi_5 \approx 0 \ .
\end{equation}

We wish to use this as a constraint to construct an action in the same
manner.  We must again consider that the constraint $\phi$ have
vanishing Dirac bracket with itself.  We find
\begin{eqnarray}
K & \equiv & \frac{i}2\{\phi,\phi\}_{D} = \frac12(p-A)^2 + \frac12
(m+\varphi)^2  \nonumber \\
& & -\, \frac{i}2\xi^\mu\xi^\nu F_{\mu\nu} + i \xi_5\xi^\mu\partial_\mu
\varphi \approx 0 \ .
\end{eqnarray}
where $F_{\mu\nu} = \partial_\mu A_\nu - \partial_\nu A_\mu$.  The
Jacobi identity again insures that there are no further constraints.  

As before, we implement these constraints by use of Lagrange multipliers, a
commuting one, $e$, and an anti-commuting one, $\lambda$,
\begin{equation}\label{eq:fullAction}
S  =  \int d\tau\, \Big[p_\mu \dot x^\mu + \frac{i}2(\xi_\mu \dot\xi^\mu
 + \xi_5  \dot \xi_5) + i\frac{\lambda}{m}\phi - eK\Big]  .
\end{equation}
We note that the action for a spinless particle can be obtained by taking
the spin variables to zero: $\xi_\mu \rightarrow 0$, $\xi_5 \rightarrow 0$.

Because we are interested only in the Thomas precession here, from now
on we consider the action with a scalar potential only, so we set
$A_\mu = 0$. Eliminating first $p_\mu$, and then $e$, in the action
Eq.~(\ref{eq:fullAction}) with $A_\mu = 0$, we find
\begin{eqnarray}\label{eq:nonlinScalar}
S & = & \int d\tau \,\Big[-\Big(m + \varphi + {i\xi_5\xi^\mu\partial_\mu
\varphi\over m + \varphi}\Big)\sqrt{-\dot x^2}\phantom{XXXXXXX} \nonumber  \\ 
& & \phantom{\int d\tau \,\Big[ } + \frac{i}2(\xi_\mu\dot\xi^\mu + \xi_5\dot\xi_5) +
i\lambda\left(\xi_5 + u\cdot\xi \right)\Big] .
\end{eqnarray}
This action is the same as the one analyzed by Martemyanov and
Shchepkin \cite{Martemyanov:1988ue}.  We note that the
Thomas-Bargmann-Michel-Telegdi equations of motion
\cite{frenkel,thomas,bargmann1959} for the spin can be found from an
analysis
\cite{Berezin,Ravndal:jq,Galvao:cu,Gomis:kh,Martemyanov:1988ue} of the
action (\ref{eq:fullAction}) with $\varphi=0$ and $A_\mu \ne 0$.

The equations of motion from the action (\ref{eq:nonlinScalar}) are
\begin{eqnarray}
\dot p_\mu & = & F_\mu , \\
\dot \xi_\mu & = &  \lambda u_\mu + {\xi_5 F_\mu \over m+\varphi} , \\
\dot \xi_5 & = & \lambda - {\xi\cdot F\over m+\varphi} ,
\end{eqnarray}
where
\begin{eqnarray}
F_\mu & = & - \partial_\mu \left(\varphi + {i\xi_5\xi\cdot\partial \varphi
\over m+\varphi}\right) \sqrt{-\dot x^2} , \\
p_\mu & = & {\partial L \over \partial \dot x^\mu} \nonumber \\
& = & (m + \varphi) u_\mu - {i\over m+\varphi} \xi_5 \xi^\nu F_\nu
{u_\mu\over\sqrt{-\dot x^2}}  \nonumber \\
& & +\, i{\lambda\over \sqrt{-\dot x^2}} P_{\mu\nu}\xi^\nu , 
\end{eqnarray}
and we have used the convenient notation
\begin{equation}
P_{\mu\nu}  =  \eta_{\mu\nu} + u_\mu u_\nu 
\end{equation}
for the projection operator perpendicular to the four-velocity.

In order to clarify the algebra in the rest of this section, we follow
Martemyanov and Shchepkin \cite{Martemyanov:1988ue} and work to lowest
order in the fermionic variables.  In this approximation we have
\begin{eqnarray}
\dot u_\mu & = & {P_{\mu\nu} F^\nu \over m + \varphi} , \\
\dot \xi_\mu & = & \left(\lambda - {u\cdot F \over m+\varphi}\right)u_\mu +
\dot u_\mu \xi_5  , \\
\dot \xi_5 & = & \lambda - {\xi_5 u\cdot F\over m+\varphi} -\dot u \cdot \xi .
\end{eqnarray}

The momentum and angular momentum of the system can be found by Noether's
theorem.  We make an infinitesimal Poincar\'e transformation of the variables
\begin{eqnarray}
\delta x^\mu & = & a^\mu + \omega^\mu{}_{\nu} \, x^\nu ,
\nonumber \\
\delta \xi^\mu & = & \omega^\mu{}_{\nu} \, \xi^\nu ,
\end{eqnarray}
and extract the conserved quantities from
\begin{eqnarray}\label{eq:noether1}
\delta S & = & \Delta\left({\partial L \over \partial \dot x^\mu} \delta
x^\mu\right) + \Delta\left({\partial^R L \over \partial \dot
\xi^\mu} \delta\xi^\mu \right) \nonumber \\
& = & a^\mu \Delta p_\mu + \frac12 \omega^{\mu\nu} \Delta J_{\nu\mu} ,
\end{eqnarray}
where $\partial^R/\partial \dot\xi^\mu$ denotes the derivative acting
from the right, and $\Delta$ denotes the difference in values between
final and initial times. In Eq.~(\ref{eq:noether1}) we have also used the
equations of motion.

We find that the total angular momentum is a sum of orbital and spin pieces
\begin{equation}
J_{\mu\nu} = L_{\mu\nu} + S_{\mu\nu} = x_{[\mu} p_{\nu]} - i \xi_\mu \xi_\nu .
\end{equation}
The total angular momentum, as well as each piece separately, obeys the Dirac
brackets relation
\begin{eqnarray}
\left\{ J_{\mu\nu}, J_{\mu^\prime\nu^\prime} \right\}_D & = &
-\, \eta_{\mu\nu^\prime} J_{\nu\mu^\prime} - \eta_{\nu\mu^\prime}
J_{\mu\nu^\prime}  \phantom{XXX} \nonumber \\ 
& & \phantom{-} +\, \eta_{\nu\nu^\prime} J_{\mu\mu^\prime}
+ \eta_{\mu\mu^\prime} J_{\nu\nu^\prime} .
\end{eqnarray}

The Pauli-Lubanski vector, 
\begin{equation}\label{eq:PL}
s_\mu = -\,\frac12 \epsilon_{\mu\nu\alpha\beta} u^\nu S^{\alpha\beta},
\end{equation}
represents the spin of the particle and is purely spatial in the rest frame
of the particle;
\begin{equation}
u\cdot s = 0 .
\end{equation}
We use the convention that $\epsilon_{0123} = +1$.
Using the identity
\begin{equation}
\epsilon_{\alpha\beta\gamma\delta} \epsilon^{\mu\nu\rho\delta} = -
\delta\strut_{[\alpha}^\mu\delta\strut^\nu_\beta\delta\strut^\rho_{\gamma]} ,
\end{equation}
we may revert Eq.~(\ref{eq:PL}) to find
\begin{equation}\label{eq:xixis}
i\xi^\alpha\xi^\beta = -\epsilon^{\alpha\beta\gamma\delta} u_\gamma
s_\delta + i(u^\alpha \xi^\beta - u^\beta \xi^\alpha)(u\cdot \xi) .
\end{equation}
Using Eq.~(\ref{eq:xixis}), we find the rate of change of $s_\mu$
\begin{eqnarray}\label{eq:sdot0}
\dot s_\mu & = & \frac{i}2 \epsilon_{\mu\nu\alpha\beta} \dot u^\nu
\xi^\alpha \xi^\beta + i \epsilon_{\mu\nu\alpha\beta} u^\nu \dot \xi^\alpha
\xi^\beta , \nonumber \\
& = & u_\mu (\dot u\cdot s) + i \epsilon_{\mu\nu\alpha\beta} u^\nu \dot
u^\alpha (u\cdot\xi) \xi^\beta  \nonumber \\ 
& & +\, i \epsilon_{\mu\nu\alpha\beta} u^\nu \dot\xi^\alpha\xi^\beta .
\end{eqnarray}
We observe that the equation of motion for $\xi^\mu$ must have the form
\begin{equation}\label{eq:needxidot}
\dot\xi^\mu = -\dot u^\mu (u\cdot \xi) + u^\mu ({\rm anything}) ,
\end{equation}
in order for the Pauli-Lubanski vector to be Fermi-Walker transported along
the worldline of the particle. That is, for $s_\mu$ to obey
\begin{equation}\label{eq:sdot1}
\dot s_\mu = u_\mu \dot u_\nu s^\nu = (u_\mu \dot u_\nu - \dot u_\mu u_\nu )
s^\nu .
\end{equation}
Equation~(\ref{eq:sdot1}) is the condition that there is no torque on
the spin. The spin thus undergoes Thomas precession, as we will see in
the next section.

\section{Thomas Precession}\label{sec:thomas}

In this section we demonstrate that a vector that undergoes Fermi-Walker
transport in a circular orbit will precess in its rest frame at the Thomas
frequency.

The spin vector of a gyroscope moved along a spacetime path $x^\mu(\tau)$
in the absence of net torque undergoes Fermi-Walker transport.  We take
laboratory time to be the worldline parameter; $\tau=t$. The rate of
change of its spin vector then is
\begin{equation}\label{eq:FW}
{ds^\mu \over dt} = \Omega^{\mu}{}_{\nu}\, s^\nu ,
\end{equation}
with 
\begin{equation}\label{eq:Omega}
\Omega^{\mu}{}_\nu = u^\mu \dot u_\nu - \dot u^\mu u_\nu  ,
\end{equation}
where the $u^\mu$ is the four velocity tangent to $x^\mu(t)$ and dot means
derivative with respect to $t$.

We make the 3+1 identifications
\begin{eqnarray}
u^0 & = & \gamma , \nonumber \\
{\bf u} & = & \gamma\, {\bf v} ,
\end{eqnarray}
and we note that the spin vector in its (non-inertial) rest
frame is
\begin{equation}
s^{\mu^\prime}_0 = \Lambda^{\mu^\prime}{}_{\nu} \, s^\nu,
\end{equation}
where $\Lambda^{\mu^\prime}{}_{\nu}$ is the Lorentz
transformation to the rest frame of the particle
\begin{equation}
\Lambda^{\mu^\prime}{}_{\nu} = \delta^{\mu^\prime}{}_{\nu} +
\pmatrix{ \gamma -1     &        -\gamma{\bf v}     \cr
         -\gamma{\bf v} &  {(\gamma - 1)\over v^2}{\bf v}{\bf v} \cr
        } .
\end{equation} 
The equation of motion satisfied by the rest-frame spin vector is
\begin{eqnarray}\label{eq:restThomas0}
{ds_0 \over dt} & = & {d\over dt}\left(\Lambda \, s\right)  =  \dot{\Lambda}\,
s + \Lambda \dot{s} \nonumber \\ 
& = & \dot{\Lambda}\,
\Lambda^{-1}\, s_0 + \Lambda\Omega\Lambda^{-1} \, s_0 .
\end{eqnarray}
The rotation matrix (\ref{eq:Omega}) is 
\begin{equation}
\Omega^{\mu}{}_{\nu} = \gamma^2 \pmatrix{ 0 &  \dot{\bf v} \cr
\dot{\bf v} &{\bf v}\dot{\bf v} - \dot{\bf v}{\bf v}\cr} .
\end{equation}
Simplifying the right hand side of Eq.~(\ref{eq:restThomas0}), we find
\begin{equation}
{ds_0 \over dt} = {\gamma -1\over v^2}
\pmatrix{ 0 & 0                                       \cr
          0 & {\bf v}\dot{\bf v} - \dot{\bf v}{\bf v} \cr } s_0 .
\end{equation}
Since the rest frame spin, $s_0$, has no time component, we have
\begin{equation}\label{eq:s0dot}
{d{\bf s}_0 \over dt} = {\gamma -1 \over v^2} \left({\bf v}\dot{\bf v} -
\dot{\bf v}{\bf v}\right)\cdot {\bf s}_0 = -\, {\gamma -1 \over v^2}
\left({\bf v}\times\dot{\bf v}\right)\times {\bf s}_0 .
\end{equation}
The acceleration of a particle in uniform circular motion with angular
velocity ${\bm \omega}$ is
\begin{equation}
\dot{\bf v} = {\bm\omega} \times {\bf v} .
\end{equation}
In the case of uniform circular motion, Eq.~(\ref{eq:s0dot}) becomes
\begin{equation}
{d{\bf s}_0 \over dt} = -\, {(\gamma -1)}\, {\bm\omega}\times {\bf s}_0 =
{\bm\Omega}_T \times {\bf s}_0 ,
\end{equation}
where ${\bm\Omega}_T$ is the Thomas frequency.

\section{String with one fixed and one massive end}\label{sec:fixedstring}

\subsection{Spinless quark}

A string with one fixed end and a massive quark on the other end is
described by an action that is the sum of the free massive point
particle action and a free string action, which we take in Polyakov
\cite{Polyakov:rd} form,
\begin{eqnarray}\label{eq:spinlessaction}
S & = & -\,{T\over 2}\int d\tau\, \int_0^1 d\sigma\,  \sqrt{-h}\,
h^{ab}\partial_a X^\mu \partial_b X_\mu \nonumber \\
& & \phantom{X} -\, m\int d\tau\, \sqrt{-\dot x^2} .
\end{eqnarray}
Here $X^\mu(\sigma,\tau)$ are the coordinates of the string
worldsheet parametrized by $\tau$ and $\sigma$, $h_{ab}$ is the
metric on the string worldsheet with $h=\det(h_{ab})$, $x^\mu(\tau)$
are the coordinates of the quark worldline, $T$ is the string
tension, and $m$ is the quark mass.  We use small latin letters for
worldsheet tensor indices.

We require that the string end at $\sigma=0$ is fixed at the origin, ${\bf
X}(0,\tau) = {\bf 0}$.  To make this an interacting theory, we must impose
the condition that the end at $\sigma=1$ ends on the quark:
\begin{equation}\label{eq:XBC}
X^\mu(1,\tau) = x^\mu(\tau) .
\end{equation}

The variation of the action under variations that preserve the
endpoint conditions, 
\begin{eqnarray}\label{eq:Xvar}
\delta {X}^\mu(0,\tau) & = &  0 , \\
\delta X^\mu(1,\tau) & = & \delta x^\mu(\tau), 
\end{eqnarray}
is
\begin{eqnarray}
\delta S & = & \int d\tau \Big( m
{\dot x^\mu \delta \dot x_\mu \over \sqrt{-\dot x^2}} 
\Big)  \nonumber \\
&  & - T\int\, d\sigma\, d\tau\, \sqrt{-h}\, h^{ab}\,\partial_a(\delta
X^\mu)\partial_b X_\mu  , \nonumber \\
& = & \int d\tau\, \delta x^\mu \Big[ {\,\mathfrak
p}^1_\mu\Big|_{\sigma=1} - {d\over d\tau} \Big({m \dot x_\mu \over
\sqrt{-\dot x^2}}\Big)\Big] \nonumber \\
\label{eq:force0}
& & \phantom{X} -  \int d\sigma\,d\tau\, \delta
X^\mu\, \partial_a  {\,\mathfrak p}^a_\mu ,
\end{eqnarray}
after an integration by parts.  Here we have used the notation
${\mathfrak p}^a_\mu$ for the current density of spacetime momentum on
the worldsheet,
\begin{equation}
{\,\mathfrak p}^a_\mu = \delta S/\delta (\partial_a X^\mu) =
-\,T \sqrt{-h}\, h^{ab}\, \partial_b X_\mu.
\end{equation}
We see in Eq.~(\ref{eq:force0}) that the force that moves the quark
arises from the boundary condition (\ref{eq:XBC}).  The key idea of our
work is to make a parallel construction with fermionic variables in
the case of a spinning quark.  In our construction, the motion of the
quark's spin comes about as a result of introducing new fermionic
variables on the string and the boundary conditions imposed upon them.

\subsection{Spinning quark} \label{sec:spinningquark}

In this section we make an ansatz for the form of the action.  In order to
have pure Thomas precession, we need an action for the fermionic variables
$\xi^\mu$ whose variation has the form
\begin{equation}
\delta S \propto \int d\tau\, \delta\xi_\mu\left[i\dot\xi^\mu + i\dot
u^\mu(u\cdot\xi)\right] ,
\end{equation}
so that we obtain Eq.~(\ref{eq:needxidot}), the condition necessary for
Thomas precession.

The term $i\delta\xi\cdot\dot u(u\cdot\xi)$ looks like $-i
\delta\xi\cdot F \xi_5/m = -i\delta \xi \cdot {\mathfrak
p}^{1} \xi_5/m$, if we use the equations of motion \hbox{$u\cdot\xi =
-\xi_5$} and make the identification $m\dot u^\mu = F^\mu$. 

We can obtain such a boundary variation by introducing worldsheet
fermionic variables $\Xi^\mu(\sigma,\tau)$ and $\Xi_5(\sigma,\tau)$ whose
boundary conditions are
\begin{eqnarray}\label{eq:XIBC}
\Xi^\mu(1,\tau) & = & \xi^\mu(\tau) ,  \nonumber \\
\Xi_5(1,\tau) & = & \xi_5(\tau) ,
\end{eqnarray}
and then replacing $\partial_a X^\mu$ in the string action
(\ref{eq:spinlessaction}) by
\begin{equation}\label{eq:modform}
\Pi^\mu_a \equiv \partial_a X^\mu - \alpha
{i\over m}\, \partial_a \Xi^\mu\, \Xi_5 
- \beta {i\over m}\, \Xi^\mu\, \partial_a \Xi_5 .
\end{equation}
We will fix the parameters $\alpha$ and $\beta$ by requiring consistency of
the equations of motion and pure Thomas precession of the spin.

In analogy to the spinless case, we take our action to be the sum of the free
Berezin-Marinov \cite{Berezin} action (\ref{eq:freefermion}) for the
particle and a Polyakov action modified by the replacement of
$\partial_aX^\mu$ by $\Pi_a^\mu$ defined in Eq.~(\ref{eq:modform}):
\begin{eqnarray}\label{eq:BM}
S & = & \int d\tau\, \Big[- m \sqrt{- \dot x^2} +
\frac{i}2(\xi_\mu \dot\xi^\mu + \xi_5\dot \xi_5)  
%% \nonumber \\
%%
%% & & \phantom{\int d\tau\, \Big[- m \sqrt{- \dot x^2}} 
%%
+\, i \lambda
(\xi_5 + u\cdot\xi) \Big]  \nonumber \\
& & -\,{T\over 2} \int d\tau\, \int_0^1 d\sigma\, \sqrt{-h}\,
h^{ab}\,\Pi_a^\mu\,\Pi_{b\,\mu}  \ .
\end{eqnarray}

\subsection{Equations of motion}\label{sec:eom}

Under variations of the $X^\mu$ and $x^\mu$ that obey the boundary
conditions Eqs.~(\ref{eq:Xvar}), we find the variation of the action to be
\begin{eqnarray}
\delta S & = & \int d\tau \Bigg[ m
{\dot x^\mu \delta \dot x_\mu \over \sqrt{-\dot x^2}} +
i\lambda\left( {\delta \dot x^\mu P_{\mu\nu}\xi^\nu\over \sqrt{-\dot
x^2}}\right)\Bigg]  \nonumber \\
&  &  -\, T\int\, d\sigma\, d\tau\, \sqrt{-h} h^{ab}\partial_a(\delta
X^\mu)\,\Pi_{b\,\mu}  ,\nonumber \\
& = &\int d\tau\, \delta x^\mu \Bigg[ {\,\mathfrak p}^1_\mu
\Big|_{\sigma=1}
% + \nonumber \\
% %
% & & \phantom{-\, \int d\tau\, \delta x^\mu \Bigg[} 
% %
\kern -4pt - {d\over d\tau} \Big({m \dot x_\mu
\over \sqrt{-\dot x^2}} + i\lambda {P_{\mu\nu}\xi^\nu\over \sqrt{-\dot
x^2}}\Big)
 \Bigg] \nonumber \\ 
& & - \int d\sigma\,d\tau\, \delta
X^\mu\, \partial_a {\,\mathfrak p}^a_\mu  ,
\end{eqnarray}
where we have again used the notation ${\mathfrak p}^a_\mu$ for the  current
density of spacetime momentum on the worldsheet, which in this case is
\begin{equation}
{\,\mathfrak p}^a_\mu = \delta S/\delta
(\partial_a X^\mu) = - T\sqrt{-h}\, h^{ab}\, \Pi_{b\,\mu} . 
\end{equation}
The vanishing of the variation $\delta S$ leads to equations of motion for
the quark and the string,
\begin{eqnarray}\label{eq:pdot}
{dp_\mu \over d\tau} & = & F_\mu = - T\, \sqrt{-h}\,
h^{1b}\,\Pi_{b\,\mu}\Big|_{\sigma=1}, \\
\label{eq:xmustring}
0 & = & \partial_a \Big( \sqrt{-h}\, h^{ab}\, \Pi_{b\,\mu}\Big),
\end{eqnarray}
where the quark's momentum is given by
\begin{equation}\label{eq:ppoint}
p_\mu  =  m u_\mu + i \lambda {P_{\mu\nu}\xi^\nu\over \sqrt{-\dot x^2}} ,
\end{equation}
with the usual projector, $P_{\mu\nu} = \eta_{\mu\nu} + u_\mu u_\nu$. 

Under variations of the fermionic variables $\Xi^\mu(\sigma,\tau)$,
$\Xi_5(\sigma,\tau)$, $\xi^\mu(\tau)$ and $\xi_5(\tau)$, obeying
\begin{eqnarray}\label{eq:xivar}
\delta \Xi^\mu(1,\tau) & = & \delta\xi^\mu(\tau) ,  \nonumber \\
\delta \Xi_5(1,\tau) & = & \delta\xi_5(\tau) ,
\end{eqnarray}
that preserve the boundary conditions (\ref{eq:XIBC}), we find the
variation of the action to be 
\begin{widetext}
\begin{eqnarray}
\delta S 
& = & \int d\tau \Big[ i\lambda\, \delta\xi^\mu\, u_\mu + i \lambda\,
\delta\xi_5 + {i\over2}\Big(\delta \xi^\mu\, \dot\xi_\mu + \xi^\mu\,
\delta\dot\xi_\mu + \delta\xi_5\, \dot\xi_5 + \xi_5\, \delta\dot\xi_5\Big)
\Big]  \nonumber \\*
&&\phantom{X}
-\,{i\over m}\int\, d\sigma\, d\tau\, 
\Big[\alpha\big(\partial_a \delta\Xi^\mu\,\Xi_5 +
\partial_a\Xi^\mu\,\delta\Xi_5\big) +\,
\beta\big(\delta\Xi^\mu\,\partial_a\Xi_5  
+ \Xi^\mu\,\partial_a\delta\Xi_5\big)\Big]{\,\mathfrak p}^a_{\mu}  , \nonumber 
\\
%% \end{eqnarray}
%% \begin{eqnarray}
%
\label{eq:var1}
& = & \int d\tau\,\Big[  \Big(i\lambda\, u_\mu - i \dot\xi_\mu + \alpha
{i\over m}\xi_5\, F_\mu \Big)\, \delta\xi^\mu + \Big( i\lambda - i\dot\xi_5 
- \beta{i\over m}\xi^\mu\,F_\mu \Big)\,\delta \xi_5 \Big] 
+\,{i\over m} \int d\tau\,\Big[ \Big(\alpha\, \delta\Xi^\mu\,\Xi_5 +
\beta\,\Xi^\mu\,\delta \Xi_5
\Big)\,{\,\mathfrak p}^1_\mu\Big]\Big|_{\sigma=0}\nonumber \\
&& \phantom{X} -\,{i\over m} \int d\sigma\,d\tau\,\Bigg\{ \Big[\alpha\,\Xi_5\,
\partial_a {\,\mathfrak p}^a_\mu + (\alpha-\beta)\,\partial_a\Xi_5 {\,\mathfrak p}^a_\mu\Big] \,\delta \Xi^\mu + \,\Big[\beta\, \Xi^\mu\,\partial_a {\,\mathfrak p}^a_\mu -\,(\alpha-\beta)\,\partial_a\Xi^\mu{\,\mathfrak p}^a_\mu\Big]\,\delta
\Xi_5 \Bigg\}.
\end{eqnarray}
\end{widetext}
Using the notation of Eq.~(\ref{eq:pdot}), and the equation of motion
(\ref{eq:xmustring}), we find the equations of motion
\begin{eqnarray}\label{eq:xieom}
\dot\xi^\mu & = & \lambda u^\mu + \alpha \xi_5 {F^\mu\over m} , \\
\label{eq:xi5eom}
\dot\xi_5 & = & \lambda - \beta \xi_\mu {F^\mu\over m} , \\
\label{eq:ximustring}
0 & = & (\alpha-\beta)\,\sqrt{-h}\, h^{ab}\Pi_{a\,\mu}\partial_b\Xi_5, \\
\label{eq:xi5string}
0 & = & (\alpha-\beta)\,\sqrt{-h}\,
h^{ab}\Pi_{a\,\mu}\partial_b\Xi^\mu.
\end{eqnarray}
These last two equations of motion, Eqs.\ (\ref{eq:ximustring}) and
(\ref{eq:xi5string}), would be automatically satisfied if $\alpha=\beta$.

The equation of motion for the metric $h_{ab}$ yields the vanishing of the 
stress-energy tensor, also known as the Virasoro constraint,
\begin{equation}\label{eq:Virasoro}
T_{ab} = \Pi_a^\mu \Pi_{b\,\mu} - \frac12 h_{ab}\,h^{cd}\,\Pi_c^\mu
\Pi_{d\,\mu} = 0 .
\end{equation}
Variation of the multiplier $\lambda$ yields the equation of motion,
\begin{equation}\label{eq:dirac}
u^\mu \xi_\mu + \xi_5 = 0 ,
\end{equation}
that becomes the Dirac equation constraint in canonical language
\begin{equation}\label{eq:lambdadirac}
p_\mu\xi^\mu + m \xi_5 \approx 0.
\end{equation}
The Klein-Gordon mass-shell condition, 
\begin{equation}\label{eq:massshell}
\frac12(p^2 + m^2) \approx 0,
\end{equation}
arises directly from squaring the momentum (\ref{eq:ppoint}).  Equation
(\ref{eq:massshell}) can also be found by taking the Dirac bracket of the
constraint (\ref{eq:lambdadirac}) with itself, as in Eq.~(\ref{eq:freeKG}).

We also need boundary conditions on the string fermionic variables at
the fixed end, $\Xi^\mu(0,\tau)$ and $\Xi_5(0,\tau)$, in order to make
the second integral in Eq.~(\ref{eq:var1}) vanish.  We cannot impose
$0=T\sqrt{-h}\,h^{1b}\Pi_{b\,\mu}|_{\sigma=0}$ because that is the
force on the fixed end, which cannot vanish.  The correct boundary
conditions are Dirichlet, of which the simplest are
\begin{eqnarray}
\Xi^\mu(0,\tau) & = & 0 , \\
\Xi_5(0,\tau) & = & 0 .
\end{eqnarray}

\section{Determination of $\alpha$ and $\beta$}\label{sec:alphabeta}

\subsection{Conservation of the Dirac equation constraint}

We begin by looking at the equations of motion in $\lambda=0$ gauge in
order to make the ideas clearer.  With $\lambda=0$, Eq.~(\ref{eq:pdot})
becomes
\begin{equation}
\dot u^\mu = {F^\mu \over m} .
\end{equation}
Using this, we simplify Eq.~(\ref{eq:xieom}) and Eq.~(\ref{eq:xi5eom}) to
\begin{eqnarray}
\label{eq:ximudot}
\dot\xi^\mu & = & \alpha\xi_5 \dot u^\mu , \\
\label{eq:xi5dot}
\dot\xi_5 & = &  - \beta\xi_\mu \dot u^\mu . 
\end{eqnarray}

The equation of motion (\ref{eq:dirac}) that leads to the Dirac equation,
\begin{equation}
u^\mu \xi_\mu + \xi_5 = 0 ,
\end{equation}
must be constant in time for consistency.  We find
\begin{eqnarray}
{d\over d\tau}\left( u^\mu \xi_\mu + \xi_5 \right)
& = & \dot u \cdot \xi + u \cdot \dot \xi + \dot \xi_5 \nonumber \\
& = & \dot u \cdot \xi + \alpha \xi_5 u\cdot\dot u -
\beta \xi\cdot \dot u \nonumber \\
& \equiv &
(1-\beta) \dot u \cdot \xi = 0.
\end{eqnarray}
Thus, for consistency we must have $\beta=1$. 

In a general gauge with $\lambda \ne 0$, we obtain a similar result:
\begin{eqnarray}\label{eq:diracdot}
{d\over d\tau}\left( p^\mu \xi_\mu + m\xi_5 \right)
& = & \alpha \xi_5 {p\cdot F\over m} + (1-\beta) {F \cdot \xi \over m}
\nonumber\\
& = & (1-\beta){ F \cdot \xi\over m} = 0,
\end{eqnarray}
as long as $p\cdot F = 0$, which is required for the consistency of the
mass-shell relation (\ref{eq:massshell}).  We take up this issue at the end
of this section.

\subsection{Thomas Precession } 

Using the equation of motion (\ref{eq:dirac}) in (\ref{eq:ximudot}), we
find
\begin{equation}\label{eq:alphaxidot}
\dot\xi^\mu = - \alpha \dot u^\mu(u\cdot \xi) + \lambda u^\mu .
\end{equation}
The analysis of Sec.~\ref{sec:pseudo} showed that it was necessary for
Eq.~(\ref{eq:needxidot}) to hold in order have pure Thomas precession.
Comparing Eq.~(\ref{eq:alphaxidot}) to (\ref{eq:needxidot}), we find
it necessary that $\alpha=1$ in order to have pure Thomas precession.

\subsection{Consistent action and boundary conditions}

Because $\alpha = \beta = 1$ from the consistency and pure Thomas
precession requirements, the string variable $\Pi_a^\mu$ is a total
derivative,
\begin{equation}
  \Pi_a^\mu =  \partial_a {\cal X}^\mu ,
\end{equation}
with 
\begin{equation}\label{eq:calXdef}
{\cal X}^\mu \equiv X^\mu - \frac{i}{m} \Xi^\mu\, \Xi_5 .
\end{equation}
Remarkably, this combination is also the key to simplifying potential
interactions of two fermions \cite{VanAlstine}.

The consistent action for a QCD string with a spinning quark on one end
that undergoes pure Thomas precession can be written using
(\ref{eq:calXdef}) as
\begin{eqnarray}\label{eq:theaction}
S & = & \int d\tau\, \Big[- m \sqrt{- \dot x^2} + \frac{i}2(\xi_\mu
\dot\xi^\mu + \xi_5\dot \xi_5) +\, i \lambda (\xi_5 + u\cdot\xi) \Big]
\nonumber \\
& & 
-\,{T\over 2} \int_0^1 d\sigma\, \int d\tau\, \sqrt{-h}\,
h^{ab}\,\partial_a{\cal X}^\mu\,\partial_{b}{\cal X}_{\mu}  .
\end{eqnarray}

Because we have $\alpha = \beta$, the equations of motion
(\ref{eq:ximustring}) and (\ref{eq:xi5string}) are automatically satisfied
and the boundary conditions on $\Xi^\mu$ and $\Xi_5$ at the fixed end can be
relaxed slightly;
\begin{equation}
\Xi^\mu(0,\tau)\, \Xi_5(0,\tau)  =  0 .
\end{equation}

\subsection{Conservation of the mass-shell constraint}

We used the condition $p\cdot F = 0$ in Eq.~(\ref{eq:diracdot}). This
condition is also necessary for the conservation of the mass-shell relation
(\ref{eq:massshell});
\begin{equation}\label{eq:pdotF}
0 = \frac12{d\over d\tau}(p^2 + m^2) = p\cdot F .
\end{equation}
We show that Eq.\ (\ref{eq:pdotF}) follows from the equations of motion of the
full action (\ref{eq:theaction}).  To begin, we use the equations of motion
(\ref{eq:pdot}), (\ref{eq:xieom}), and (\ref{eq:xi5eom}) and the expression
(\ref{eq:ppoint}) for the quark's momentum to calculate the boundary value
\begin{eqnarray}
m\dot{\cal X}^\mu\Big|_{\sigma=1} & = & m\dot x^\mu - i\dot\xi^\mu \xi_5 - i
\xi^\mu\dot\xi_5 \nonumber \\ 
& = & m\dot x^\mu - i\Big(\lambda u^\mu + \frac{\xi_5}mF^\mu\Big)\xi_5
\nonumber \\
& & -i\xi^\mu\Big(\lambda - \frac{\xi\cdot F}m\Big) \nonumber \\
& = & \sqrt{-\dot x^2}\,p^\mu + \frac{i}m\xi^\mu\xi\cdot F .
\end{eqnarray}
Using the nilpotency of $\xi\cdot F$ and the Virasoro constraint
(\ref{eq:Virasoro}),  we find 
\begin{eqnarray}
p\cdot F &=& {m\over\sqrt{-\dot x^2}}\partial_0 {\cal
X}^\mu\Big|_{\sigma=1}F_\mu \nonumber \\
& = & \Big[{m\sqrt{-h} h^{1b}\over\sqrt{-\dot x^2}} \partial_0 {\cal X}^\mu
\partial_b{\cal X}_\mu\Big]\Big|_{\sigma=1} \nonumber \\
& = & \Big[{m\sqrt{-h} h^{1b}\over 2\sqrt{-\dot x^2}}
h_{0b}\,h^{cd}\partial_c {\cal X}^\mu \partial_d{\cal
X}_\mu\Big]\Big|_{\sigma=1} \nonumber \\
& = & \delta^1_0 \Big[{m\sqrt{-h}\over 2\sqrt{-\dot x^2}}
\,h^{cd}\partial_c {\cal X}^\mu \partial_d{\cal
X}_\mu\Big]\Big|_{\sigma=1}\nonumber \\
& = & 0 .
\end{eqnarray}
We have used $h^{1b}h_{0b} = \delta^1_0 = 0$ in the last line.  If we do
the same analysis keeping $\alpha$ and $\beta$ arbitrary, after a bit of
algebra we find
\begin{equation}
p\cdot F = i(1-\beta){\lambda\over \sqrt{-\dot x^2}}(\xi\cdot F),
\end{equation}
again showing the necessity of having $\beta=1$.

\section{Fermionic Gauge Invariance}\label{sec:fermiongauge}

The string portion of the action (\ref{eq:theaction}) has two fermionic
constraints,  
\begin{eqnarray}\label{eq:stringHsym}
\Phi_\mu & = & \Pi_\mu - \frac{i}{m}P_\mu\Xi_5  \approx  0 ,\\
\Phi_5 & = & \Pi_5 + \frac{i}{m}P_\mu\Xi^\mu  \approx 0 ,
\end{eqnarray}
where $\Pi_\mu$, $\Pi_5$, and $P_\mu$ are the momenta conjugate to
$\Xi^\mu$, $\Xi_5$, and $X^\mu$ respectively.  The fermionic
constraints together with the Virasoro constraints (\ref{eq:Virasoro})
are all first-class.  It is easy to compute the Poisson brackets
\begin{eqnarray}
\{\Phi_\mu,\Phi_\nu\} = \{\Phi_\mu,\Phi_5\} = \{\Phi_5,\Phi_5\} = 0.
\end{eqnarray}
Because the stress tensor (\ref{eq:Virasoro}) is traceless, there are
only two independent Virasoro constraints, which we may take in the
form \cite{Green:sp}
\begin{equation}
  L_{\pm} = \frac12\left(P \pm T {\cal X}^\prime\right)^2.
\end{equation}
After a bit of algebra, we find the Poisson brackets
\begin{eqnarray}
  \{ L_{\pm}(\sigma), L_{\pm}(\varrho)\} & = &
     T\left(L_{\pm}(\sigma) +
     L_{\pm}(\varrho)\right)\delta^\prime(\sigma- \varrho) ,  \nonumber \\
  \{L_{\pm}(\sigma),L_{\mp}(\varrho)\} & = &0 , \nonumber \\
  \{L_{\pm}(\sigma),\Phi_\mu(\varrho)\} & = & 0 , \nonumber \\
  \{L_{\pm}(\sigma),\Phi_5(\varrho)\} & = & 0 . 
\end{eqnarray}
By acting in combination on the fields $A$ through Poisson brackets,
\begin{equation}
\delta_H A = \{A, H^\mu \Phi_\mu + H_5 \Phi_5\} , 
\end{equation}
the constraints (\ref{eq:stringHsym}) generate the following fermionic
gauge invariance of the action
\begin{eqnarray}
\delta_H X^\mu &=& \frac{i}{m}\Big(H^\mu\Xi_5 + \Xi^\mu H_5\Big) , \nonumber \\
\delta_H \Xi^\mu &=& H^\mu , \nonumber \\
\delta_H \Xi_5 &=& H_5 .
\end{eqnarray}
Here $H^\mu = H^\mu(\sigma,\tau)$ and $H_5=H_5(\sigma,\tau)$ are
Grassmann-valued functions on the string worldsheet.  This is not an
invariance of the particle action, so the gauge parameters $H$ must
vanish at the boundary.  Obviously ${\cal X}^\mu$ is gauge invariant,
\begin{equation}
\delta_H {\cal X}^\mu  \equiv 0,
\end{equation}
so the string action (\ref{eq:theaction}) is invariant as well.
Because we have as many first-class constraints as fermionic
variables, there are no dynamical fermionic degrees of freedom on the
string; except for their values at the boundary, they are pure gauge.  

Just as for a free Dirac particle action, the particle piece of the
action (\ref{eq:BM}) has a local supersymmetry generated by the Dirac
constraint
\begin{equation}
 \phi = p\cdot \xi + m \xi_5 \approx 0 .
\end{equation}
The gauge variation of $x^\mu$ is 
\begin{equation}
 \delta_\eta x^\mu = \{x^\mu,i\eta\phi\}_D = i\eta\xi^\mu .
\end{equation}
The gauge variations of the other variables are
\begin{eqnarray}
  \delta_\eta p_\mu & = & 0 , \nonumber \\
  \delta_\eta \xi^\mu & = & - \eta p^\mu , \nonumber  \\ 
  \delta_\eta \xi_5 & = & - \eta m .
\end{eqnarray}
The Lagrange multiplier fields $\lambda$ and $e$ have gauge variations
\begin{eqnarray}
  \delta_\eta \lambda & = & - \dot \eta m , \nonumber  \\ 
  \delta_\eta e & = & - \,{2 i \lambda \eta \over m} .
\end{eqnarray}

\section{Energy and Angular Momentum}\label{sec:noether}

The numerical quantization of the relativistic flux tube model starts
from the conserved quantities of the system. The canonical form of
these quantities for the string with a spinning quark are only slightly
different from those for the spinless case.  In this section we
calculate the four-momentum and the angular momentum of our system.

The action, Eq.~(\ref{eq:theaction}), is invariant under infinitesimal
translations and Lorentz transformations,
\begin{eqnarray}
\delta x^\mu(\tau) & = & a^\mu + \omega^{\mu}{}_{\nu}
x^\nu(\tau),\nonumber \\
\delta X^\mu(\sigma,\tau)&=&a^\mu+\omega^{\mu}{}_{\nu}
X^\nu(\sigma,\tau),\nonumber \\ 
\delta \xi^\mu(\tau) & = & \omega^{\mu}{}_{\nu} \xi^\nu(\tau) ,
\nonumber  \\
\delta \Xi^\mu(\sigma,\tau) & = & \omega^{\mu}{}_{\nu} \Xi^\nu(\sigma,\tau) .
\end{eqnarray}
Noether's theorem guarantees the existence of conserved total momentum
${\cal P}_\mu$ and conserved total angular momentum ${\cal J}_{\mu\nu}$,
which can be computed from the vanishing change in the action,
\begin{equation}\label{eq:defpJ}
\delta S = a^\mu \Delta {\cal P}_\mu  + \frac12 \omega^{\mu\nu}
\Delta {\cal J}_{\nu\mu} = 0,
\end{equation}
assuming the use of the equations of motion.
From Eq.~(\ref{eq:defpJ}), we find explicitly
\begin{eqnarray}\label{eq:p}
{\cal P}^\mu & = & {p^\mu} + \int_0^1 d\sigma \, {P^\mu} \\
\label{eq:J}
{\cal J}_{\mu\nu} & = & {x_{[\mu}}p_{\nu]} - {\xi}_{[\mu} \pi_{\nu]}
+ \int_0^1 d\sigma \, \Big({X_{[\mu}}P_{\nu]} - {\Xi_{[\mu}}\Pi_{\nu]} \Big)
\nonumber \\
& = & {x_{[\mu}}p_{\nu]} - i\xi_{{\mu}} \xi_{\nu} +
\int_0^1 d\sigma \,{{\cal X}_{[\mu}}P_{\nu]} ,
\end{eqnarray}
where ${\cal X}^\nu = X^\nu - \frac{i}{m}\Xi^{\nu}\Xi_5$.  In the last
line of Eq.~(\ref{eq:J}) we have used the constraints (\ref{eq:Chis}) and
(\ref{eq:stringHsym}).

\section{Discussion}\label{sec:discussion}

Starting from the requirement that an action for a spinning quark on
the end of a string should lead to pure Thomas precession of the quark
spin, we have shown how to construct a consistent action for a massive
spinning quark on the end of a QCD string whose other end is fixed.
To do so, we introduced additional fermionic variables on the QCD
string itself and required that the equations of motion of the quark's
fermionic variables arise from boundary conditions on the string.  The
two parameters we introduced, $\alpha$ and $\beta$, are fixed to unity
by the requirements of pure Thomas precession and consistency of the
equations of motion, respectively.  

Our action has a consistent fermionic symmetry that allows us to gauge
away the stringy fermionic degrees of freedom.  Other authors
\cite{Horava,Polyakov:1997tj,Vyas:2000qi} have argued for the
existence of a supersymmetry on the QCD string worldsheet when there
are spinning quarks on the boundary.  It would be interesting to know
if our fermionic symmetry is that supersymmetry. As one piece of
evidence, we can make contact with the Wilson super-loop approach
\cite{Polyakov:1997tj,Migdal:1996rn,Vyas:2000qi} by making a change of
variables in our action, Eq.~(\ref{eq:theaction}).  When we make the
variable change
\begin{eqnarray}
  X^\mu & \rightarrow & X^\mu + \frac{i}{m} \Xi^\mu \Xi_5, \nonumber \\
  x^\mu & \rightarrow & x^\mu + \frac{i}{m} \xi^\mu \xi_5,
\end{eqnarray}
in the string and quark actions respectively, we are led to the
Polyakov bosonic string action plus a quark action
\begin{equation}\label{eq:modquark}
  S_q = \int d\tau\,\left(-m\sqrt{-\dot x^2} - i u^\mu \dot u^\nu
  \xi_\mu\xi_\nu + \cdots \right),
\end{equation}
where we have neglected to write the kinetic terms for the fermions,
the Dirac constraint, some higher-order fermionic pieces, and a total
time derivative.  The second term yields the interaction of the
quark's spin with the string worldsheet.  Unfortunately, this action
is not simple and does not seem to have a reasonable canonical
formulation, unlike Eq.~(\ref{eq:theaction}).  We also
note that the second term of Eq.~(\ref{eq:modquark}) is similar to one
added by the Martemyanov and Shchepkin \cite{Martemyanov:1988ue},
though there are additional terms in (\ref{eq:modquark}) not present
in their action.

We have not considered the case of a quark at each end, however the
generalization is immediate.  We introduce a set of Grassmann
variables for each quark, $\xi^\mu_i$, $\xi_{5\, i}$, with $i=1,2$. In
this case, however, the variables for one quark commute with those of
the other \cite{VanAlstine}, just as the gamma matrices of two
different fermions commute.  We also introduce a set of worldsheet
fermionic variables for each quark and make the generalization in the
string action
\begin{equation}
  X^\mu \rightarrow {\cal X}^\mu = X^\mu - i \sum_{i=1,2}
  \frac{1}{m_i} \Xi_i^\mu \Xi_{5\, i} .
\end{equation}
Though the $\Xi$ variables of each quark are Grassmann-valued, the
$\Xi$ variables of one quark should commute with the $\Xi$ variables
of the other.  

We have not considered the case of a massless quark, which appears to
be somewhat problematic in our formalism.  On the other hand, there is
no problem treating very light, but still massive, quarks in this
formalism.

Although we have partially analyzed our action in the general case of
$\alpha \neq \beta$, we have not pursued the analysis with $\alpha
\neq 1$ because we are most interested in the phenomenologically
relevant $\alpha = \beta = 1$ case.  In the more general case, the
fermionic constraints on the worldsheet are not all first-class and
some of the fermionic variables on the worldsheet may become
dynamical, though additional terms in the action may be necessary to
preserve the first-class nature of the Virasoro constraints.

We hope to present soon a numerical quantization of this action.

\section*{ACKNOWLEDGMENTS}
We thank C. Goebel for reading the manuscript.  This work was
supported in part by the US Department of Energy under Contract
No.~DE-FG02-95ER40896.

\appendix

\section{Pseudoclassical Mechanical Conventions}\label{sec:PCMC}

We take a canonical form for an action to have the  velocities to the
right of the momenta,
\begin{equation}
  S = \int L\, d\tau =\int d\tau \, \left[p_i \dot q^i + \pi_\alpha \dot
  \xi^\alpha - H(q,p,\pi,\xi)\right] ,
\end{equation}
where $q^i$ and $p_i$ are bosonic variables and $\xi^\alpha$ and $\pi_\alpha$
are fermionic variables and $H$ is a Grassmann even function. The
variation of $H$ under a change of a fermionic variable such as
$\delta\xi^\alpha$ is
\begin{equation}
  \delta H = {\partial^R H \over \partial \xi^\alpha}\delta\xi^\alpha ,
\end{equation}
where ${\partial^R H /\partial \xi^\alpha}$ denotes the derivative from the
right.  We could equally well have used 
\begin{equation}
  \delta H = \delta\xi^\alpha{\partial^L H \over \partial \xi^\alpha} ,
\end{equation}
since it has the same value.  Variation of the action leads to the
usual canonical equations of motion,
\begin{eqnarray}\label{eq:Acan}
  \dot q^i & = & {\partial H \over \partial p_i} \nonumber \\
  \dot p_i & = & -\,{\partial H \over \partial q^i} \nonumber \\
  \dot \xi^\alpha & = & -\,{\partial^R H \over \partial \pi_\alpha} 
  = {\partial^L  H \over \partial \pi_\alpha} \nonumber \\
  \dot \pi_\alpha & = & -\,{\partial^R H \over \partial \xi^\alpha} 
  = {\partial^L H \over \partial \xi^\alpha} .
\end{eqnarray}
The last two relations of Eq.~(\ref{eq:Acan}) follow because $H$ is an
even Grassmann parity function.  These relations can be succinctly
summarized by the introduction of a Poisson bracket
\begin{equation}
  \dot z = \{ z, H\},
\end{equation}
where $z$ is any of $q^i$, $p_i$, $\xi^\alpha$, or $\pi_\alpha$ and
the Poisson bracket of any two functions $A$ and $B$ is
\begin{eqnarray}
  \{A, B\} & = & \sum_i \left({\partial A \over \partial q^i}{\partial
  B \over \partial p_i} - {\partial A \over \partial p_i}{\partial B
  \over \partial q^i} \right) \nonumber \\*
& & + \sum_\alpha \left({\partial^R A \over \partial
  \xi^\alpha}{\partial^L B \over \partial \pi_\alpha} + {\partial^R A
  \over \partial \pi_\alpha}{\partial^L B \over \partial \xi^\alpha}
  \right) .
\end{eqnarray}

\section{Dirac brackets}\label{sec:DiracBrackets}

When a system has second-class constraints and one wishes to set them
strongly to zero, consistency requires that the Poisson brackets
of the system be modified so that the Poisson bracket of any
second-class constraint with any other phase space function is
identically zero.  This modified Poisson bracket is called a Dirac
bracket.  

The simplest example is illuminating, though artificial. We imagine a
dynamical system in which there are $2N$ phase space variables and two
second-class constraints, $q_N \approx 0$ and $p_N \approx 0$.  These
constraints hold for all time so $q_N$ and $p_N$ are irrelevant
variables; no physical quantity should depend upon them.  The correct
procedure is to ignore these variables and replace the Poisson bracket,
\begin{equation}
    \{A, B\} = \sum_{i=1}^N \left({\partial A \over \partial
  q^i}{\partial B \over \partial p_i} - {\partial A \over \partial
  p_i}{\partial B \over \partial q^i} \right) ,
\end{equation}
by the Dirac bracket
\begin{equation}\label{eq:ignoreN}
    \{A, B\}_D = \sum_{i=1}^{N-1} \left({\partial A \over \partial
  q^i}{\partial B \over \partial p_i} - {\partial A \over \partial
  p_i}{\partial B \over \partial q^i} \right).
\end{equation}
The Dirac bracket (\ref{eq:ignoreN}) of $q_N$ or $p_N$ with any other
phase space function will obviously vanish.

For a more complicated system with second-class constraints
$\chi_i\approx 0$, the Dirac bracket is less obvious.  The
matrix of Poisson brackets of the second-class constraints 
\begin{equation}\label{eq:chiDelta}
\{\chi_i,\chi_j\} = \Delta_{ij}, %% \qquad \det\Delta  \not\approx 0 ,
\end{equation}
has non-vanishing determinant, and is therefore invertible.  We denote
the matrix inverse to $\Delta_{ij}$ by $\Delta^{ij}$, and define the
Dirac bracket of any two functions $A$ and $B$ as
\begin{equation}
  \{A,B\}_D \equiv \{A, B\} - \{A,\chi_i\}\Delta^{ij}\{\chi_j,B\} .
\end{equation}
The desired property now follows,
\begin{eqnarray}
  \{A,\chi_k\}_D & = & \{A, \chi_k\} -
  \{A,\chi_i\}\Delta^{ij}\{\chi_j,\chi_k\}  \nonumber\\
 & = & \{A, \chi_k\} -
  \{A,\chi_i\}\Delta^{ij}\Delta_{jk} \nonumber \\
 & = & \{A, \chi_k\} -
  \{A,\chi_i\}\delta^{i}_{k}  \equiv 0.
\end{eqnarray}
We note that some authors use $\{A,B\}^*$ to denote the Dirac bracket.

\end{document}